# On Developers' Personality in Large-scale Distributed Projects: The Case of the Apache Ecosystem

Fabio Calefato, Giuseppe Iaffaldano, Filippo Lanubile
University of Bari
Italy
{fabio.calefato, giuseppe.iaffaldano, filippo.lanubile}@uniba.it

Bogdan Vasilescu
Carnegie Mellon University
USA
vasilescu@cmu.edu

## ABSTRACT

Large-scale distributed projects are typically the results of collective efforts performed by multiple developers, each one having a different personality. The study of developers' personalities has the potential of explaining their' behavior in various contexts. For example, the propensity to trust others, a critical factor to the success of global software engineering – has been found to influence positively the result of code reviews in distributed projects.

In this paper, we perform a quantitative analysis of developers' personality in open source software projects, intended as an extreme form of distributed projects in which no single organization controls the project. We mine ecosystem-level data from the code commits and email messages contributed by the developers working on the Apache Software Foundation (ASF) projects, as representative of large scale-distributed projects.

We find that developers become over time more conscientious, agreeable, and neurotic. Moreover, personality traits do not vary with their role, membership, and extent of contribution to the projects. We also find evidence that more open and more agreeable developers are more likely to become project contributors. [1]

## CCS CONCEPTS

• **Human-centered computing** → Collaborative and social computing; *Empirical studies in collaborative and social computing*

## KEYWORDS

Personality traits; large-scale distributed projects; team-building; social and human aspects; open source software; apache; ecosystems

## 1 INTRODUCTION

Large-scale distributed projects are typically the results of collective efforts performed by multiple developers, each having their different personality [39]. There are many definitions of personality as established by previous research on psychology. In general, personality is viewed as a dynamic and organized set of traits that create the unique patterns of behavior, thoughts, and feelings of a person [28]. As such, the study of personalities has the potential of explaining developers' behavior in various contexts [26].

Calefato & Lanubile [4] and Calefato et al. [5] found initial evidence that the propensity to trust – i.e., the facet of personality representing the individual disposition to perceive the others as trustworthy – is positively correlated with the chances of successfully accepting contributions in code review tasks. In this paper, we follow up that study, broadening the investigation on the effects of the various personality traits of developers', rather than only trust.

As a first step, we focus on studying the effects of developers' personality in open source software (OSS) projects, intended as an extreme form of distributed projects in which no single organization controls the project [14]. OSS project teams consist, in fact, of different types of contributors [18][19]. At the center are *core contributors*, who are part of the development team and contribute the largest portion of the code base; they also review external code contributions and provide guidance to newcomers. *Peripheral contributors*, instead, are not part of the development team and most of them do not remain involved with the project for long; they are typically involved with contributing bug fixes, adding projects documentation, and code refactoring. Finally, *one-time contributors* (OTCs) are on the very fringe of the peripheral developers since they have exactly one code contribution accepted to the project repository. According to this layered organizational structure (also known as the *onion model* [41]), developers migrate from the edges to the core of OSS projects through a gradual socialization process. Ducheneaut [13] found that contributions coming from submitters who are known to the core development team have higher chances of being accepted, as core developers also use the record of interactions as signals for judging the quality of proposed changes. Consistently, Tsay et al. [36] and Yu et al. [42] have also observed that the social distance between the contributor and the integrator largely influences the chances of accepting code contributions. With the rise of 'transparent' social-coding platform such as Bitbucket and GitHub, integrators make inferences about the quality of contributions, not only by looking at their technical quality but also using more 'social' auxiliary indicators such as developers' track record (e.g., previous contributions accepted) and reputation (e.g., number of stars and followers in GitHub) [12][21].

Socialization in large, distributed projects typically happens through written communication [13]. Because psycholinguistic research has shown that written interaction style (e.g., emails) reflects one's personality [15][29], here we mine ecosystem-level data from 32 Apache Software Foundation (ASF) projects and analyzed the emails and code commits contributed by 144 developers. We collect historical data on the communication and development activity of these developers and extract their personality profiles seeking differences with respect to their role, the degree of productivity, and project tenure. Finally, we develop a regression model for understanding the extent to which personality is related to the likelihood of becoming a project contributor – i.e., author of at least one accepted commit.

We find that developers' personality evolves over time as more conscientious, agreeable, and neurotic. Moreover, personality traits do not vary with their role, membership, and extent of contribution to the projects. We also find evidence that stronger openness and agreeableness traits scores are associated with higher chances for developers to become a project contributor. As a further contribution, we speculate that our research findings can be exploited for using personality profiles in recommending mentors to newcomers in existing team or for building new ones.

The remainder of this paper is organized as follows. In Section 2, we review related work on personality in software development. In Section 3, we present our research questions. In Section 4, we describe the experiment. Results are reported and discussed, respectively, in Section 5 and 6. Finally, we draw conclusions in Section 7.

## 2 BACKGROUND

Prior studies of personality in software development typically use *objective personality tests* to identify differences among developers [11]. Tests like the Myers-Briggs Type Indicator (MBTI) [23] and the NEO-PI [9] have been developed by psychologists under the assumption that personality is primarily conscious and can be directly accessed and measured through self-assessment questionnaires. These resources are used by researchers because they are considered reliable and easy-to-use instruments.

In a recent study, Xia et al. [39] explored the relationship between the mix of developers' personality in software teams and the extent of success of their projects. About 300 developers from 28 projects took the *Dominance, Influence, Steadiness, and Compliant* (DISC) personality test; then, their self-reported personality scores were correlated with project success scores, as evaluated by project managers. They found that that the project with the highest success scores were those executed by project teams (i) with dominant managers or (ii) with a mix of influential members and less dominant ones. The unreliability of self-reporting data, however, has been long acknowledged in psychology research (e.g., see [20])

Many studies derive personality based on a popular framework, the *Big-Five personality model* (or *Five-Factor Model*) [10], a taxonomy with two levels of traits and facets:

- *Agreeableness* trait: a person's tendency to be compassionate and cooperative toward others. Facets are trust, cooperation, altruism, sympathy, modesty, moralism.

- *Conscientiousness* trait: a person's tendency to act in an organized or thoughtful way. Facets are self-efficacy, orderliness, dutifulness, achievement-striving, self-discipline, cautiousness.

- *Extraversion* trait: a person's tendency to seek stimulation in the company of others. Facets are friendliness, gregariousness, assertiveness, activity level, excitement-seeking, cheerfulness.

- *Neuroticism* (or emotional stability) trait: the extent to which a person's emotions are sensitive to the person's environment. Facets are anxiety, anger, depression, self-consciousness, immoderation, vulnerability.

- *Openness* trait: the extent to which a person is open to experiencing a variety of activities. Facets are imagination, artistic interests, emotionality, adventurousness, intellect, liberalism.

Tausczik & Pennebaker [35] found that every trait in the Big-Five model is strongly and significantly associated with theoretically-appropriate patterns of word use, indicating strong connections between language use and personality. Consistently, other studies have successfully derived personality traits from the analysis of written text [15], such as emails [29].

Rigby & Hassan [27] studied the Big-Five personality traits of the four top developers of the Apache httpd project against a baseline built using LIWC on the entire mailing list corpus. Their preliminary results showed that **two of the developers responsible for the major Apache releases have similar personalities, which are also different from the baseline** extracted from the email corpus contributed by the other developers.

Bazelli et al. [1] performed a quasi-replication of the previous study using developer-related data collected from Stack Overflow instead of a mailing list. They found that the **top reputed authors are more extroverted compared to medium and low reputed users**, a personality profile consistent to the one observed by Rigby & Hassan [27] for the two top Apache httpd developers.

Rastogi & Nagappan [26] analyzed the development activity of about 400 active GitHub developers, investigating the relationship between the Big-Five model personality traits and productivity. They found that **developers with different levels of contributions have different personality profiles**, specifically those with high or low levels of contributions are more neurotic compared to the others. Besides, the **personality profiles of most active contributors were found to change across two consecutive years**, evolving as more conscientious, more extrovert, and less agreeable.

Calefato & Lanubile [4] and Calefato et al. [5] investigated the relationship between project success and propensity to trust. To avoid subjectivity in the assessment of project success, they approximated the overall performance of two Apache projects with the history of such successful collaborations, i.e., code review of pull requests in GitHub. They found preliminary evidence that the **propensity to trust of code reviewers (integrators) is an antecedent of pull request integration**. To avoid subjectivity in the assessment of the propensity to trust, they used the *Linguistic Inquiry and Word Count* (LIWC) psycholinguistics dictionary to analyze word usage in writing [25][35] and extracted developers' agreeableness scores based on the Big-Five model.

Overall, the findings from previous research on personality in software development suggest that personality of developers varies with the degree of contribution (e.g., between core and peripheral developers), reputation, and changes over short time spans (i.e., developers' project tenure). Because previous research relied consistently on the Big-Five model and LIWC, for the sake of comparison, in the following we also rely on the same personality framework and psycholinguistic resource.

## 3 RESEARCH QUESTIONS

The review of previous work on personality in software developed revealed several potential factors related to developers' activity and social status, which may reflect their personality. Therefore, to further our understanding of developers' personality and their effects, we focus on studying their activities in both the technical part (i.e., code development through commits) and the social part (i.e., communication through emails) of the ASF ecosystem. Building on the findings from previous work, in the following we formulate five of research questions.

With respect to project tenure, Rastogi & Nagappan [26] found that developers' personality changes over short time spans. Hence, seeking further evidence on the stability of personality traits of developers, we borrow the same research question:

*RQ1 – Does developers' personality change over short time spans?*

These changes in personality observed by Rastogi & Nagappan [26] may be due to the different type of tasks that developers perform and their responsibilities in the community. Therefore, we derive and compare the personality of developers splitting the corpus of emails before and after they gain write-access to the core repository (i.e., they become integrators who can accept and merge others' contributions), a sign that they were promoted to the core development team of a project. Accordingly, our second research hypothesis is:

*RQ2 – Does developers' personality change after becoming a core member of a project's development team?*

According to the findings of Rigby & Hassan [27] and Bazelli et al. [1], the personality of top-reputed users in software communities are different from the others. In our experimental scenario, that would suggest potential differences in the personality traits between peripheral and core developers.

*RQ3 – Does developers' personality vary with the type of contributor (i.e., peripheral vs. core developer)?*

According to Rastogi & Nagappan [26], the personality of developers varies with their degree of code contributions, too. We seek confirming evidence for this finding. Hence, our fourth research question is formulated as follows:

*RQ4 – Does personality vary with the degree of development activity?*

Finally, Calefato & Lanubile [4] and Calefato et al. [5] found that the propensity to trust of integrators who perform code review is predictive of the likelihood of accepting external code contributions. Yet, trust is one facet of the Big-Five model traits and previous research did not look at the effects of the personality of developers who author those contributions. Here, we bridge this gap. Hence, the fifth research question is:

*RQ5 – What personality traits are associated with the likelihood of becoming a contributor?*

## 4 EMPIRICAL STUDY

To answer our research questions, we analyzed the development activity of ASF projects. Specifically, we performed the analysis and regression modeling of repository data to assess quantitatively the effects of developers' personality. In the following, we first report how personality was objectively measured (Section 4.1) and describe the experimental dataset (Section 4.2); then, we detail the regression model and analysis (Section 4.3).

**Table 1: The data sources used to collect the data in our study**

| Data source | Data extracted |
|---|---|
| ASF web pages | Project name |
| | Status (active, incubating, retired) |
| | Dev. language |
| | Category |
| | Repository URI (git, svn) |
| | Mailing-list URIs (dev, user) |
| Project mailing lists | Mailing list name |
| | Emails (body, subject, sender, recipient, timestamp) |
| | Developers' email addresses |
| Git repositories | Repository (id, # commits, timestamp first and last commit) |
| | Developer's info (id, email) |
| | Commit metadata (repository, sha, author id, commiter id, timestamp, commit message, files changed, src files changed, # additions and deletions) |

## 4.1 Measuring Personality

To obtain an objective, quantitative measure of developers' personality, we relied on the *Personality Insights* API,[2] an IBM Watson service that leverages LIWC. The service uses linguistic analysis to detect three types of tones from written text: social, emotional, and writing style. Specifically, the social tone measures the social tendencies (i.e., the Big-Five high-level personality traits) in people's writing. Provided with some textual input, the API returns a JSON document with values in [0, 1] for each of the five personality traits of the writer.

## 4.2 Dataset

To build our experimental dataset, we mined several data sources. The full list of the metadata extracted from each data source is reported in Table 1. Also, the scripts developed for mining the data source are made available, along with the extracted data, on GitHub[3] for the sake of replicability.

The first data source is the **official web pages**[4] of the ASF projects. The list of projects was obtained by developing a custom web scraper, using the Python Scrapy[5] library. Some project metadata were also extracted through the scraper, namely the status of the project (i.e., *Active*, *Retired*, *Incubating*), its development language (e.g., *Java*, *C++*) and category (e.g., *database*, *web*), the mailing-list archive URIs, and the URI of its code repository. At the end of this stage, a list of 176 ASF projects was retrieved.

The second data source is the **mailing list archives**. Through the scraper, we retrieved for each project the URIs of the dev mailing list (i.e., containing development-oriented discussion such as bug reports) and user mailing list (i.e., containing general purpose discussion such as release announcements) archived in the mbox format. Then, we forked, updated and run the mlstats[6] tool to download the mailing lists to a local MySQL database. Finally, a script was developed to collect all the emails sent by a sender to a project mailing list in a month and compute the Big-Five personality scores using the Personality Insights API to analyze the aggregated body of text. Eventually, for each sender, we obtained a time series of personality scores arranged by month and per project. At the end of this step, 106 mailing lists were entirely downloaded, for a total of 1.35M emails from ~38,000 senders.

The third and last data source is the **project code repositories**. We downloaded to a local machine a clone of the repository for each project using Git. The other projects were discarded. Then, a script was written to parse the commit history of each project clone and save to the MySQL database the relevant metadata extracted, such as the IDs of the author and of the integrator, the time stamps, the file changed, the number of additions and deletions, etc. (refer to Table 1 for the full list). The number of commits is used as a proxy for project size; likewise, the delta in years between the first and the last commit is used as a proxy for its longevity. At the end of this step, we selected and cloned the Git repositories of 56 ASF projects using Git, totaling ~206K commits and 5,080 developers.

**Alias unmasking**. Looking at the extracted data, we observed that, in many cases, the same sender used multiple email addresses to post messages to project mailing lists. This aliasing issue not only affects the communication but also the project development, as developers often committed code contributions using different emails. Therefore, we applied a procedure used in Vasilescu et al. [37] to 'unmask' alias email addresses. First, for each developer/sender stored in our database, an alias set was computed and assigned a unique identifier (UID in the following). Then, we stored a hash map of these UIDs so that, whenever a database entry was processed, the map was used to replace its table ID with the associated unique UID. The map contained the UIDs of 46,304 unique developers who either sent emails or contributed code to the AFS projects. No obvious cases of mislabeling were detected during the manual analysis of a sample.

## 4.3 Analysis

We run several statistical tests using the R statistical package.

To answer RQ1 (variation of personality over time) and RQ2 (variation with project membership), we used the Wilcoxon Signed-Rank test as a non-parametric alternative to t-test for paired samples. Instead, for the research questions RQ3 (variation of personality with the type of contributor) and RQ4 (variation with the degree of contribution), we used the Wilcoxon Rank Sum (or Mann-Whitney U) test, as a non-parametric alternative to t-test for unpaired samples. We use the p-values to determine statistical significance and supplement those with effect size (Cliff's $\partial$).

---

[2] www.ibm.com/watson/services/personality-insights
[3] https://github.com/collab-uniba/personality
[4] https://projects.apache.org/projects.html
[5] https://scrapy.org
[6] https://github.com/MetricsGrimoire/MailingListStats

Finally, to answer RQ5, we fit a logistic regression model to our data to assess the likelihood for a developer to become a project contributor, using personality scores as predictive factors. The regression variables included in the model are detailed below.

*Response*. The response variable is `contributor`, a dichotomous yes/no variable indicating whether a developer has authored at least one commit successfully integrated into a project repository.

We note that here we considered only those developers who contributed code changes and exchanged a sufficient amount of text via emails, from which a personality profile could be extracted.

*Main predictors*. We include `openness`, `agreeableness`, `neuroticism`, `extraversion`, and `conscientiousness`, that is, one predictor for each of the Big-Five high-level personality traits.

*Controls*. Our control variables include: `word_count`, a proxy for the communication and social activity of the developer in the community through email messages from which personality traits are extracted; `project_size`, computed as the total number of commits in the projects; `project_age`, measured in number of years. The last two variables are intended to reflect that it may be harder for developers to start contributing to long-running projects that have a large code base. Yet, we had to discard both because, when we checked the model for collinearity issues among the predictors using the VIF (variance inflation factor), their value was close to 4.

We fit the model using the `glm` function in R. Coefficients are considered important when statistically significant at the 1-5% level ($p < 0.0.1$ or $p < 0.05$). We also performed analysis of deviance to assess the features contributing the most to predicting the response variable `contributor`. Finally, we evaluated our model's fit using the $R^2$ measure, which describes the proportion of variance explained by the model, and AUC, to assess the classification ability of our model against random guessing.

## 5 RESULTS

In the following, we report the results of our analyses, grouped by research question.

**RQ1 (time).** We compared the trait scores for those developers who participated actively over a time span of three consecutive years (2014-2016). In the following, we report the boxplots and the results of the Wilcoxon Signed-Rank tests comparing the personality of developers working during those years (see Figure 1 and Table 2, respectively). The tests revealed statistically significant differences ($p < 0.05$, with Bonferroni correction) between 2014 and 2016 for openness, agreeableness, and neuroticism. The Cliff's $\partial$ values (between -0.72 and -0.73) showed large effect sizes for these differences.

**RQ2 (membership).** For each developer with write access to a project's code repository, we retrieved the date of the first commit that they review and accept to integrate. Then, we used this date as an approximation of the moment when the developers became core team members of that project. Accordingly, we used that date to split the personality scores for of the developers into two groups, i.e. before and after becoming a project's core team member. Figure 2 shows the differences in the five personality scores between the two paired groups and Table 2 reports the results of the Wilcoxon

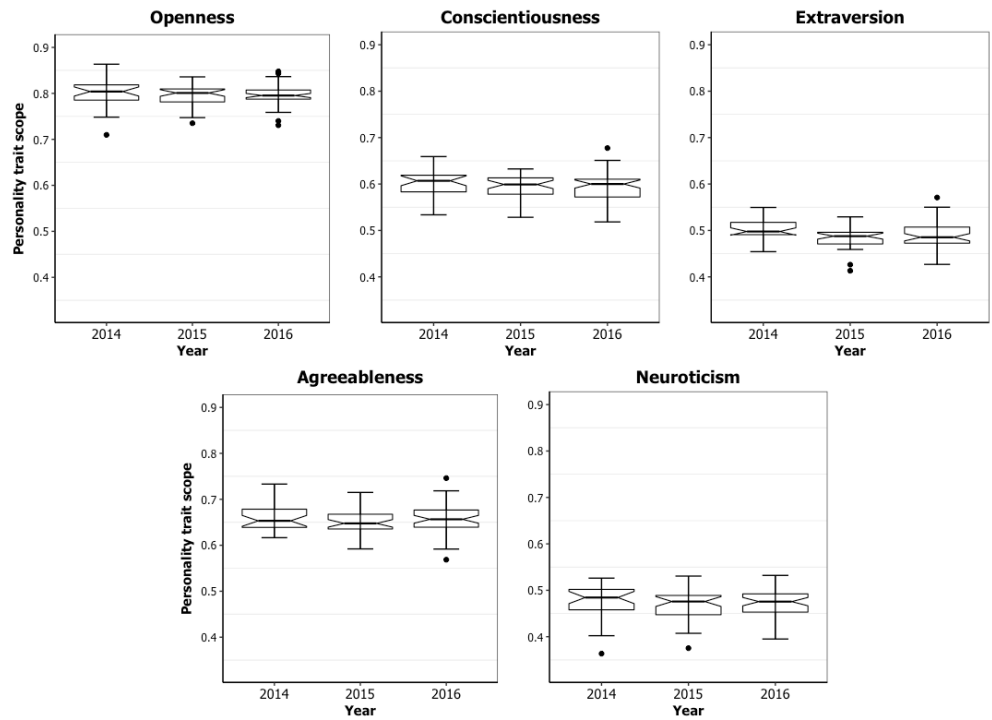

Figure 1: Differences in the personality traits of the developers participating in the years 2014-2016.

**Table 2. Results of the Signed-Rank test for the comparison of mean personality trait scores (** $p < 0.05$).**

|  | Paired groups | Trait | V | p-value | Cliff's ∂ |
|---|---|---|---|---|---|
| RQ1 | Years 2014 - 2015 | Openness | 400 | 0.25 | -0.69 |
|  |  | Conscientiousness | 400 | 0.28 | -0.69 |
|  |  | Extroversion | 500 | 1.00 | -0.68 |
|  |  | Agreeableness | 400 | 0.57 | -0.69 |
|  |  | Neuroticism | 400 | 0.33 | -0.69 |
|  | Year 2015 - 2016 | Openness | 800 | 0.57 | -0.70 |
|  |  | Conscientiousness | 800 | 0.32 | -0.70 |
|  |  | Extroversion | 800 | 0.40 | -0.70 |
|  |  | Agreeableness | 700 | 0.12 | -0.71 |
|  |  | Neuroticism | 700 | 0.17 | -0.70 |
|  | Year 2014 - 2016 | **Openness** | **700** | **0.04**** | **-0.72** |
|  |  | Conscientiousness | 700 | 0.06 | -0.72 |
|  |  | Extroversion | 700 | 0.21 | -0.72 |
|  |  | **Agreeableness** | **600** | **0.01**** | **-0.73** |
|  |  | **Neuroticism** | **600** | **0.03**** | **-0.73** |
| RQ2 | Before vs After becoming core-team members | Openness | 70 | 0.5 | 0.07 |
|  |  | Conscientiousness | 90 | 0.2 | 0.15 |
|  |  | Extroversion | 50 | 0.5 | -0.12 |
|  |  | Agreeableness | 50 | 0.5 | -0.16 |
|  |  | Neuroticism | 70 | 0.5 | 0.18 |

Signed-Rank tests between them. No significant differences were retuned by the tests.

**RQ3 (contributor type).** We separated the personality scores in two groups, *peripherals* (i.e. the commit authors without commit access to the repositories) and *core* developers (i.e., project members with access). For the sake of space, here we omit to report the boxplot. Results of the Wilcoxon Rank-Sum tests are reported in Table 3, which show a significant difference in the case of extraversion ($p < 0.5$) and agreeableness ($p < 0.01$). Yet, Cliff's ∂ estimates revealed negligible effect sizes (∂ = –0.07 and ∂ = –0.05, respectively).

**RQ4 (degree of contribution).** We took the core and peripheral groups created for RQ3, and further split them according to the level of contribution. Specifically, we found the median number of authored commits in the peripheral, and split it into two subsets, *authored-commits_high* and *authored-commits_low*; similarly, we obtained the subgroups *integrated-commits_high* and *integrated-commits_low* considering the median number of integrated commits from the core group. We then performed pairwise comparisons of the mean personality scores between the four subgroups. Results are in shown in Table 3. The Wilcoxon Rank-Sum tests revealed a few cases of statistically significant differences between the pairs. However, the Cliff's ∂, revealed small effect sizes (∂ between 0.11 and 0.18).

**RQ5 (contribution likelihood model).** In Table 4, we report the results of the logistic model. We observe that the control variable `word_count` is not statistically significant. The statistically significant predictors are `openness` ($p < 0.01$) and `agreeableness` (p < 0.05). The results of the analysis of deviance show that personality feature that contributes most (i.e., percentage of deviance explained) to predicting the likelihood of becoming a `contributor` is `agreeableness` (~61%). To evaluate the fit, we compute the $R^2$ measure for the statistical model developed, which shows that our model is capable of explaining the 14% of the variability in the data ($R^2 = 0.14$). Besides, we used the under the ROC curve (AUC) metric to measure the AUC performance of the model. The result (AUC=0.82) shows that the model performs better than the random prediction model (AUC=0.5).

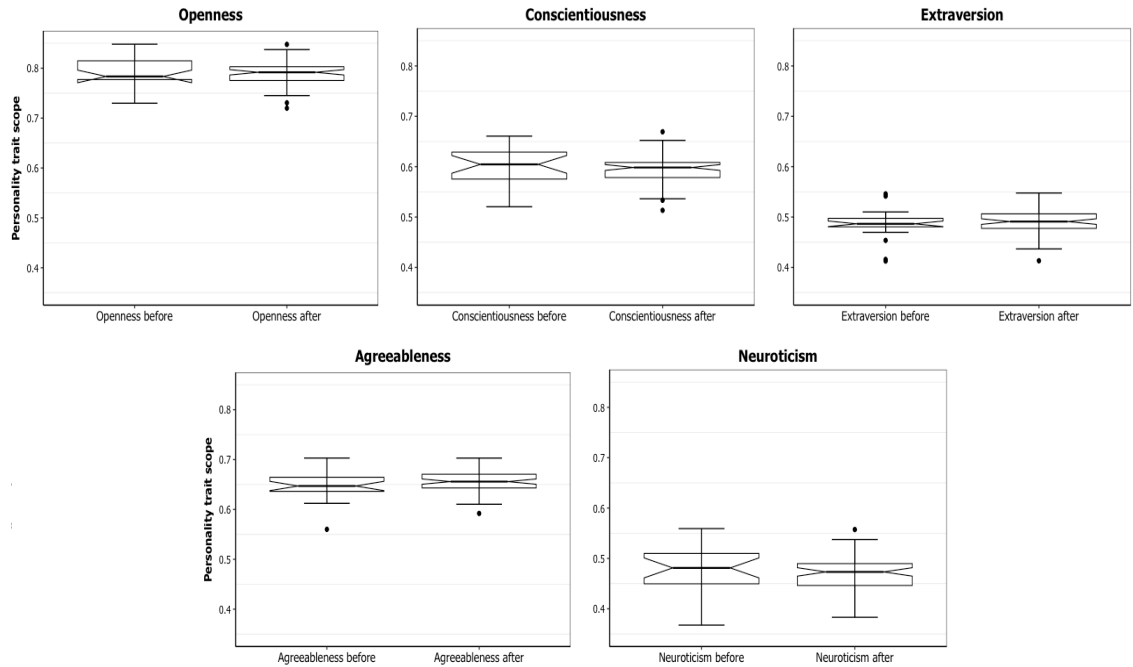

**Figure 2: Differences in the personality traits of the developers before and after becoming core team members.**

**Table 3. Results of the Wilcoxon Rank-Sum test for the unpaired comparison of mean personality trait scores between core and peripheral developers (\* p< 0.01, \*\* p < 0.05).**

|  | Paired groups | Trait | W | Cliff's ∂ |
|---|---|---|---|---|
| RQ3 | Core vs. Peripheral developers | Openness | 8e+05 | -0.02 |
|  |  | Conscientiousness | 9e+05 | -0.01 |
|  |  | Extroversion | 8e+05* | -0.07 |
|  |  | Agreeableness | 8e+05** | -0.05 |
|  |  | Neuroticism | 8e+05 | -0.02 |
| RQ4 | High vs. Low contribution (authors) | Openness | 3e+05* | -0.13 |
|  |  | Conscientiousness | 3e+05* | -0.16 |
|  |  | Extroversion | 4e+05 | -0.05 |
|  |  | Agreeableness | 3e+05* | -0.17 |
|  |  | Neuroticism | 4e+05** | 0.06 |
|  | High vs. Low contribution (integrators) | Openness | 80,000 | 0.03 |
|  |  | Conscientiousness | 90,000* | 0.11 |
|  |  | Extroversion | 80,000 | 0.06 |
|  |  | Agreeableness | 70,000* | -0.16 |
|  |  | Neuroticism | 90,000* | 0.18 |

Intuitively, the results above tell us that stronger openness and agreeableness traits scores are associated with higher chances for developers to become a project contributor. To provide a more quantitative interpretation, we note that the mean agreeableness and openness values in the dataset are, respectively, 0.65 and 0.79. Given the logistic model in Table 4, the probability of becoming a contributor for those developers with agreeableness and openness scores below averages is 62%, compared to 93% for developers with scores equal to or above averages (+31%).

## 6 DISCUSSION

The results reported in the previous section add to the body of existing evidence about mining the personality traits of developers from software-related repositories. The evidence from our experiment is somewhat in contrast with previous findings. While contrasting results may be due to the different psychometric tools adopted in the studies and their respective reliability (a potential limitation that we discuss in Section 6.1), below we speculate otherwise.

**Table 4: Logit model of contribution likelihood explained by personality traits**

|  | Estimates | Std. Error | Dev. |
|---|---|---|---|
| (Intercept) | -43.45** | 13.99 |  |
| log(word_count) | -0.10 | 0.13 | 1.55 |
| openness | 38.80** | 15.01 | 3.10 |
| agreeableness | 32.92* | 15.34 | 12.98 |
| neuroticism | -22.19 | 13.51 | 3.57 |
| extraversion | 2.49 | 12.24 | 0.01 |
| conscientiousness | 5.08 | 14.70 | 0.12 |

$R^2 = 0.14$; Num. obs. = 144
AUC = 0.82;

\*\* p < 0.01    \* p < 0.05

Regarding the study of how personality evolves in developers, with respect to RQ1 (*Does personality change over short time spans?*) we found evidence that developers evolve as more open, agreeable, and neurotic. In other words, they tend to become, respectively: (i) more capable of expressing their feelings; (ii) more cooperative, altruistic, and prone to trust the others; (iii) more prone to worry, self-conscious, and susceptible to stress. Rastogi & Nagappan [26] found that GitHub developers' personality changed over time, too. However, in contrast with our results, they found developers to evolve as more conscientious and extrovert, and less agreeable over two consecutive years. While further investigations are needed to explain these differences, we note these claims are made despite some negligible to small effect sizes (Cohen's ∂) observed. The small effect sizes may also account for the contrasting findings regarding RQ4 (*Does personality change with the degree of contribution?*) since they found that developers who contribute more score high on openness, conscientiousness, extraversion, and neuroticism, and low on agreeableness, while we observed no variations.

Our other findings also show a sort of 'stability' in developers' personality traits, which are not affected by project membership (*RQ2 – Does personality change with project membership?*) and the type of contributor (*RQ3 – Does personality change between core and peripheral developers?*). This evidence contrasts with the findings of Rigby & Hassan [27] and Bazelli et al. [1], who found that top developers have different personality traits than the others. However, Rigby & Hassan [27] reported data about 2 top developers only. The finding contrasting with Bazelli et al. [1], instead, is arguably explained by the difference in experimental domains. In fact, they analyzed the communication and question-answering activity of developers within Stack Overflow, while we looked at email and coding in the Apache ecosystem.

Finally, regarding RQ5 (*What personality traits are associated with the likelihood of becoming a contributor?*), the logistic model developed showed that the control variable `word_count` (i.e., the proxy for the social activity in the community) is not statistically significant. This means that the amount of communication that a developer exchanges in the ASF communities is not associated with the likelihood of becoming a contributor. However, previous research (e.g., [13]) has found that contributions coming from submitters who are known to the core development team have higher chances of being accepted. Combined, these findings indicate that the quality of the messages and their recipients are important to become a contributor, rather than the amount of communication exchanged. Furthermore, the result of the logistic regression also shows that more open and agreeable developers are more likely to author one or more commits that are successfully integrated into a project repository. This finding is consistent with the results of our previous work [4][5] where it was found that more agreeable integrators are more likely to accept the pull requests during code review sessions. Agreeableness, in fact, is associated with the propensity to trust other, being empathetic, and avoiding harsh confrontations – facets of personality that are 'helpful' during cooperative tasks such as code reviews, where more open/agreeable contributors and integrators are likely to collaborate with less friction. Previous research on OSS projects has highlighted that newcomers face several entry barriers, not only technical but also social, when placing their first contribution, leading in many cases to dropouts [31][33]. Hence, our finding suggests that more open/agreeable core members may be better suited to shepherd newcomers during their immigration phase (i.e., onboarding and first contributions) [2][32]. In previous work, Canfora et al. [8] successfully tested an approach to recommend the 'right mentors' among core team members to guide OSS project newcomers. Their recommendations were based on discovering previous interactions through emails on topics of shared interest. Our finding indicates a possible extension of their approach, which includes mining personality profiles to identify candidate mentors among those core-team members who are more open and agreeable.

More in general, finding the 'right mix' of personalities has potential implications regarding team-building not only for OSS projects but also for commercial ones, especially if distributed. In previous research, Yang et al. [40] found that agreeableness helped teammates coordinate through the development of shared mental models, thereby enhancing software team performance. Karn et al. [17] found in lab experiment that software teams reported higher cohesion and performance in cases of both homogeneity in personality type and some mixtures of types.

## 6.1 Limitations

Regarding the generalizability of our results, since the Apache ecosystem may not be representative of all types of large, distributed projects, especially commercial, we acknowledge the need to gather further evidence. Yet, independent replications are also welcome, as we have made all the code and the entire dataset available online.[7]

The main limitations of the study revolve around the instrument chosen to model and extract the personality profiles of the developers. First, the Big-Five model and the other similar personality frameworks are mostly used by researchers because they are considered valid and easy-to-use instruments. While this is true for the initial administration, the interpretation of the results and the analysis of their practical implications are not straightforward and require properly trained professionals [22]. Still, Cruz et al. [11] found that existing studies on the personality of software developers seldom involved psychologists, suggesting this as a possible explanation of the contradictory findings revealed by their systematic mapping study. We acknowledge the same limitation in our study, which may account for the contrasting findings between this and the previous studies on developers' personality. As future work, we intend to involve a fellow researcher with a background in psychology / psycholinguistic to help us with the use of the personality instruments and the consequent interpretation of the measurement and findings.

Second, the use of the Personality Insights service allowed us to extract a large number of developers' personality profiles from emails without relying on self-reported data. By exploiting the large amount of communication messages archived in these software-related repositories – i.e., the toolset belonging to the social-programmer ecosystem [30] – more and more recent studies like ours have started to employ natural language processing (NLP) instruments for the automatic analysis of their content (e.g., [3][6][7]). Yet, many of these tools have not been designed or trained for handling technical content typical of the software domain [24]. For instance, Jongeling et al. [16] have compared various sentiment analysis tools used in previous studies in software engineering and found that not only they disagree with the manual labeling of corpora performed by individuals, but also, they do not agree with each other. This issue raises a serious concern that researchers may draw diverging conclusions if different NLP tools are used, especially those not specifically trained for the specific purpose and lexicon. While we acknowledge this as a potential threat to instrumentation validity for our work, we note that previous research (e.g., [15]) has found evidence that the personality traits can be successfully derived from the analysis of short, written texts such as emails. We also point that we employed the Personality Insights service on emails only after parsing them to remove (most of) the technical content therein through *ad hoc* regular expressions. Finally, in similar research, Wang & Redmiles [38], used the LIWC tool (upon which Personality Insights builds) to compute the baseline trust of developers parsing the content of their emails. They compared the results against those obtained using another linguistic resource (i.e., the NRC lexicon) and found them to converge. Nonetheless, we reserve the comparison of our findings obtained with LIWC and Personality Insights API against other similar toolsets as future work.

Finally, we note that in our analyses we have considered someone to be a developer if (s)he has committed changes to the project's source code repository, regardless of the type of change. However, not all commits necessarily include changes to source code files (e.g., documentation). In our future work, we will look at the files touched in the commits to include only those containing changes to source files.

## 6.2 Research Agenda

---

[7] https://github.com/collab-uniba/personality

While our investigation provides some new insights on the effects of personality in software development, there are still many unanswered research questions. Below, we present a couple of them, which we intend to answer in future work.

**Combined effect of personalities in collaboration**. In our current and previous work [4][5], we have analyzed the effect of personality traits on the likelihood of accepting contributions. Yet, we analyzed the personality of contributors and integrators in isolation. This may explain, for example, why the regression model developed for RQ5 fits our data marginally (it explains the 14% of the variability), indicating the existence of other explanatory factors not included in our current model. As future work, we intend to analyze the combined effects of developers' personalities during joint tasks such as code review.

**Effectiveness in mentor recommendation and team building**. Earlier, we argued the potential benefits of leveraging personality profiles in recommending mentors for newcomers in existing team or in building new teams. Yet, we ignore whether similar personalities should be matched or instead, a select mixture of personality types must be sought. We also ignore the effect of the task to perform. As future work, we intend to test how effective the envisioned solutions would be compared to the current state of the art.

## 7 CONCLUSIONS

In this paper, we presented a quantitative analysis of the personality traits of the developers working in the Apache ecosystem. Developers' personalities were extracted from the projects' mailing list archives and modeled on the Big-Five personality framework, using the IBM Personality Insights service.

We found that developers' personality evolves over time as more conscientious, agreeable, and neurotic. Moreover, personality traits do not vary with their role, membership, and extent of contribution to the projects. We also developed a regression model and found that the openness and agreeableness traits are antecedents of successfully becoming project contributor. This finding has practical implications in recommending the right mentors to project newcomers as well as for building new teams by considering the analysis of personalities for the prospect team members.

Our findings are in contrast with previous work on the personality of developers, thus calling for further studies and replications.